\documentclass[useAMS]{mn2e}
\usepackage{psfig}

\title[X-Ray Sources and Star Clusters]{Displacement of X-Ray Sources
from Star Clusters in Starburst Galaxies} 

\author[P. Kaaret et al.]{P.~Kaaret$^1$, A.~Alonso-Herrero$^{2,3}$,
J.S.~Gallagher, III$^4$, \newauthor G.~Fabbiano$^1$, A.~Zezas$^1$,
M.J.~Rieke$^2$ \\ $^1$Harvard-Smithsonian Center for Astrophysics, 60
Garden St., Cambridge, MA 02138, USA\\ $^2$Steward Observatory,
University of Arizona, Tucson, AZ 85721,  USA\\ $^3$Departamento de
Astrof\'{\i}sica Molecular e  Infrarroja, IEM, Consejo Superior de
Investigaciones Cient\'{\i}ficas, 28006  Madrid, Spain\\ $^4$Department
of Astronomy, University of Wisconsin at Madison, 475 North Charter
Street, Madison, WI 53706, USA}

\date{Accepted . Received  ; in original form }

\pagerange{\pageref{firstpage}--\pageref{lastpage}}
\pubyear{2004}

\begin{document}

\maketitle

\label{firstpage}

\begin{abstract}

We examine the spatial offsets between X-ray point sources and star
clusters in three starburst galaxies.  We find that the X-ray sources
are preferentially located near the star clusters.  Because the star
clusters are very good tracers of the star formation activity in the
galaxies, this indicates that the X-ray sources are young objects
associated with current star formation.   We find significant
displacements of the X-ray sources from the clusters.  These
displacements are likely due to motion of the X-ray sources and
indicates that they are X-ray binaries.  We find that brighter X-ray
sources preferentially occur closer to clusters.  The absence of
very bright sources at large displacements from clusters may help
constrain models of the sources.

\end{abstract}


\section{Introduction}
\vspace{-0.1in} 

Starburst galaxies contain prominent populations of X-ray sources
\cite{fabbiano89} including unusually luminous ones
\cite{kaaret01,zezas02}.  The young stellar ages within starbursts
suggest that the associated X-ray sources are likely to be young
objects \cite{kilgard02}.  Strong X-ray variability in many sources
further suggests those sources are accreting compact objects
\cite{makishima00,kaaret01,fabbiano03}.

In starburst galaxies, a substantial fraction of young stars are found
in compact, luminous star clusters \cite{meurer95}.  Indeed, Tremonti
et al.\ (2001) and Harris et al.\ (2001) find evidence that most high
mass stars in starbursts could form in dense clusters which quickly
dissolve to feed the surrounding field.  Consistent with this, the
young compact cluster R136a contains a rich population of massive
stars, evidently following a standard Salpeter-like upper initial mass
function \cite{massey98}.  However, there also are indications that
some compact young massive clusters have either a flatter than normal
upper mass function or a cutoff at low mass
\cite{sternberg98,smith01,mccrady03}.  The current data are
insufficient to distinguish between these two possibilities
\cite{mengel02}.  In either case, the clusters would contain relatively
more massive stars, and thus potentially more massive X-ray binaries. 

Star clusters in starburst regions usually are centrally concentrated,
with as much as $10^6 M_{\odot}$ of stars within a half light radius of
a few parsecs (e.g.\ Ho \& Filippenko 1996a,b).  Stellar encounters in
such dense clusters may enhance the production of X-ray binaries
\cite{portegies99}. Interactions of binaries in clusters
\cite{phinney91} and binary recoil following supernovae leading to
compact object formation \cite{nelemans99} can lead to ejection of
X-ray binaries from their point of origin.  Therefore, if the X-ray
sources are X-ray binaries, then they may be expected to be spread over
a larger spatial area than the star clusters.

Here, we study the spatial offsets between X-ray point sources and star
clusters in three starburst galaxies with dense clusters and deep,
high-resolution X-ray observations.  We describe our sample, the
observations, and our analysis in \S~2, the results in \S~3, and
discuss the implications in \S~4.

\section{Data}


\begin{figure*} 
\centerline{\psfig{file=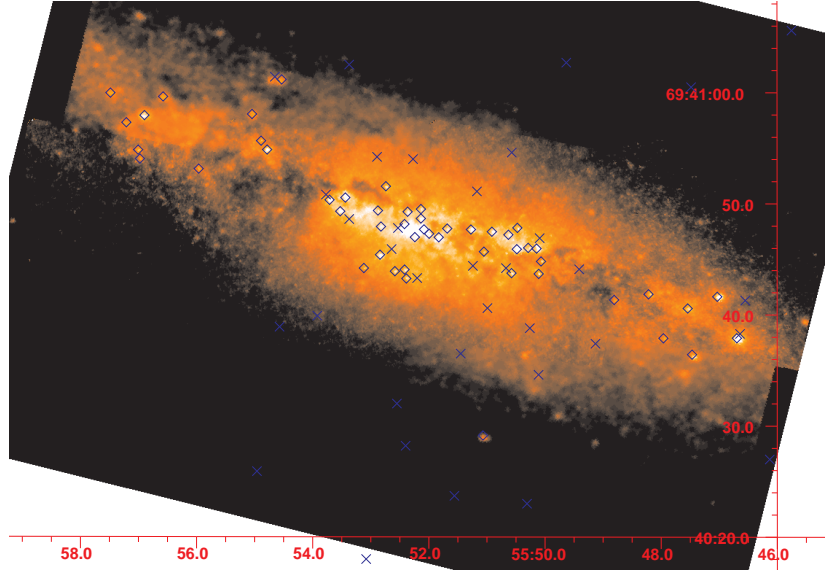,height=3.0in}} 
\caption{False color infrared image of M82.  The NIC2 data were used
for the inner parts of the mosaic and the NIC3 for the remainder. 
Candidate star clusters are marked as diamonds.  X-ray sources are
marked as X's.  There are 8 sources in the sample which do not appear
on the figure.  North is up, and the axes are R.A. and DEC in J2000
coordinates.} \label{m82ir} \end{figure*}

Our selected galaxies are: M82, NGC 1569, and NGC 5253.  All three are
relatively nearby, within 4~Mpc.  Hence, relatively dim X-ray sources
can be detected with modest exposures.  NGC 5253 is a dwarf galaxy
which has been undergoing a starburst for no more than 10~Myr
\cite{beck96} and the young star clusters have ages between 1~Myr and
8~Myr \cite{calzetti97,tremonti01}.  M82 is the prototypical starburst
galaxy.  The nuclear star burst has been active for roughly 10~Myr
\cite{satyapal97} and may consist of two events with ages of 5~Myr and
10~Myr \cite{mccrady03}.  There is also a more spatially extended
``fossil'' starburst which occurred 0.6--1~Gyr ago
\cite{degrijs01,degrijs03}.  NGC 1569 is a dwarf irregular galaxy which
has been in a starburst phase for 10-20~Myr \cite{hunter00}.

\subsection{Star clusters}

For each galaxy, we require the positions of the star clusters. For NGC
1569, we used the list of star clusters compiled by Hunter et al.\
(2000) based on HST observations in the optical band.  Because this
galaxy has a relatively low gas/dust content and therefore low internal
extinction, the optical data should provide a complete cluster list.

M82 is a larger galaxy and has high obscuration near its core, where
most of the star clusters are located \cite{mcleod93}.  Comparison of
optical and IR images shows that many clusters are missed in the
optical due to the high obscuration.  We chose to derive a cluster list
from IR observations obtained with NICMOS on HST.  The observations and
reduction of data to images are described in Alonso-Herrero et al.\
(2003). We used a sliding cell algorithm to detect compact sources
within  the images from the NIC2 camera using the F160W filter with a
central wavelength 1.6~$\mu$m and covering a wavelength range of
1.4-1.8~$\mu$m, and the NIC3 camera using the F166N filter with a
central wavelength 1.66~$\mu$m and a 1\% bandpass (images in the F160W
filter were not available with NIC3).  The NIC2 images are higher
resolution (75~mas pixels) and cover the inner regions of the galaxy. 
The NIC3 images extend the imaged field at lower resolution (200~mas
pixels).

We found clusters in each image.  Clusters found in multiple images
were used to align the relative positions of the various NICMOS
images.  We determined the absolute astrometry of the NICMOS images
using the positions of four clusters detected in the 2MASS survey which
lie well outside the crowded central region of the galaxy.  After
applying one global shift to the NICMOS coordinates, the positions of
all four clusters agree within $0.5\arcsec$ with the 2MASS source
positions.  A mosaic of the NICMOS data is shown in Fig.~\ref{m82ir}.

For NGC 5253, we also used NICMOS observations taken using the NIC2
camera and the F160W filter.  The details of these observations, the
analysis, and the cluster list will be presented elsewhere
(Alonso-Herrero et al., in preparation).  The NICMOS field of view is
$19\arcsec \times 19\arcsec$ and covers the central region of the
galaxy.  The astrometry was set using the accurate position for the
dominant 1.3~cm radio source \cite{turner00} which is coincident with
one of the IR clusters.  Due to the small field of view, the roll angle
uncertainty from the HST aspect solution leads to only small
uncertainties in the cluster positions.

\subsection{X-ray sources}

We extracted the longest available observation of each galaxy from the
Chandra X-Ray Observatory data archive.  All observations employed the
Advanced CCD Imaging Spectrometer and the High-Resolution Mirror
Assembly.  The data for M82 \cite{griffiths00} and NGC 1569
\cite{martin02} have been published previously.  We reanalyzed the data
to have consistent results for all three galaxies.

For each galaxy, we extracted an image of X-rays in the 0.3--8~keV
band.  We calculated an exposure map for an assumed powerlaw spectrum
with interstellar absorption.  The powerlaw photon index was fixed to
1.8 in all cases, and the absorption column density set to the Galactic
line-of-sight value for each galaxy.  We detected sources using the
routine {\it wavdetect} in CIAO version 2.3 requiring a source
significance of $3\sigma$.  We compared the resulting source list with
optical images from the POSS2 to remove foreground stars and with the
NED extragalactic object database to remove background AGN.  We used
X-ray sources with either AGN or optical star counterparts (from the
USNO B catalog) to check the astrometry.  The absolute X-ray source
positions should be accurate to  better than $1\arcsec$.

Because spectral fitting is not feasible except for the few brightest
sources, we calculated an X-ray flux using the same spectrum assumed
for the exposure map and correcting only for interstellar absorption
within our Galaxy.  If the sources have intrinsic absorption or there
is absorption within the host galaxy, the true fluxes may be higher. We
calculated a luminosity for each X-ray source assuming that it is at
the distance to the corresponding galaxy which we take to be 3.6~Mpc
for M82, 2.2~Mpc for NGC 1569, and 3.3~Mpc for NGC 5253
\cite{gibson00}.  Note that we calculate luminosities assuming
isotropic emission.  If the X-rays are beamed, then the luminosities
would be lower.  The luminosities are calculated for the 0.3--8~keV
band.  The weakest detected sources are $3 \times 10^{36} \rm \, erg \,
s^{-1}$ for M82, $5 \times 10^{35} \rm \, erg \, s^{-1}$ for NGC 1569,
and $1 \times 10^{36} \rm \, erg \, s^{-1}$ for NGC 5253.

While some of the X-ray sources may be supernova remnants (SNRs) or
young supernovae, the fraction of such sources is likely to be small. 
Martin et al.\ (2002) identify only one Chandra X-ray source in NGC
1569 as a possible SNR.  Griffiths et al.\ (2000) report three close
coincidences (within $1\arcsec$) of Chandra X-ray sources with radio
SNRs in M82.  However, the sources have hard X-ray spectra rather the
soft spectra which are found for SNRs \cite{prestwich03} and young
supernovae \cite{kaaret01b}.


\begin{figure} \psfig{file=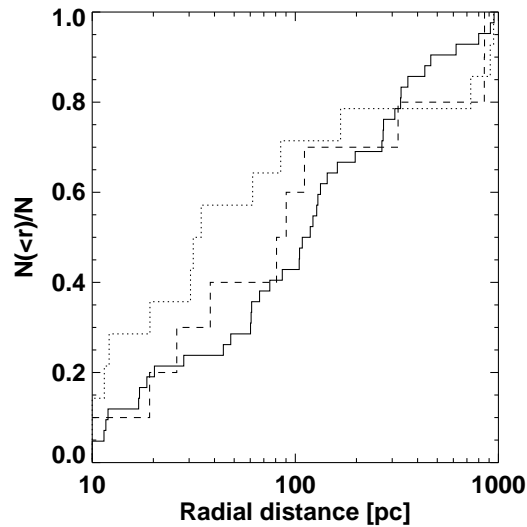,width=3.0in} \caption{Cumulative
distribution of X-ray source displacements for sources in the
luminosity range $10^{36} {\rm \, erg \, s^{-1}} < L_X <  10^{38} {\rm
\, erg \, s^{-1}}$. The solid line is for M82, the dotted line for NGC
1569, and the dashed line for NGC 5253.  Each histogram is normalized
by the number of X-ray sources within 1~kpc of a star cluster for that
galaxy.} \label{dhist} \end{figure}

\begin{table}
\begin{center}
\caption{Statistics of the Displacement Distributions.}\label{disttable}
\begin{tabular}{lcccc}
\hline
Galaxy   & X-ray    & Clusters  & \multicolumn{2}{c}{Displacement (pc)}  \\
         & Sources  &           & Average        & Median \\ \hline
M82      & 42       & 50        & 201 $\pm$  37  & 118$^{+11}_{-14}$ \\
NGC 1569 & 14       & 58        & 218 $\pm$  95  &  34$^{+54}_{-4}$ \\
NGC 5253 & 10       & 13        & 239 $\pm$ 106  &  90$^{+21}_{-52}$ \\ \hline
\end{tabular}
\end{center}
\noindent The Table contains for each galaxy: the galaxy name, the
number of X-ray sources within 1 kpc of a star cluster, the number
of star clusters, and the average and median of the displacements
of X-ray sources from star clusters.
\end{table}

\section{Results}

To investigate the relation between star clusters and X-ray sources, we
found the nearest star cluster to each X-ray source.  For X-ray sources
born in clusters, this provides a lower bound on the displacement of
the source from its parent cluster.  The uncertainty in the relative
alignment of the X-ray source and star cluster positions is $1\arcsec$
or less, corresponding to uncertainties in the displacements of about
10~pc.  Displacements smaller than 10~pc have been set to 10~pc. 
Displacements smaller than about 20~pc should be considered as upper
bounds.  We examine only sources within 1~kpc of a star cluster.  The
distributions of the spatial displacements are shown in
Fig.~\ref{dhist} and some statistics of the distributions are shown in
Table~\ref{disttable}.

Our cluster lists contains all of the most luminous clusters for each
galaxy, but may be incomplete at low luminosities.  To determine if our
results are sensitive to the completeness of the sample, we varied the
detection threshold for the clusters in M82 and found the distributions
of the spatial displacements of X-ray sources from the clusters for the
various cluster lists.  We found no significant changes in the
distribution for reasonable changes in the threshold.  This may be
because the lower luminosity clusters tend to be located near brighter
clusters already in the list.  Thus, the displacements are decreased
only slightly.

To determine if the clustering of the X-ray sources near the star
clusters is statistically significant, we generated random sets of
uniformly distributed sources and found the spatial displacements from
the star clusters following the same procedures used for the actual
X-ray sources.  The displacement distributions of the X-ray sources are
inconsistent with that expected for a uniform distribution of sources. 
The probabilities of chance occurrence from a uniform distribution are
$2 \times 10^{-6}$ for M82, $5 \times 10^{-5}$ for NGC 1569, and $2
\times 10^{-4}$ for NGC 5253.  For each galaxy, the average
displacement of X-ray sources from star clusters is significantly
smaller than for the random source distribution, indicating that the
X-ray sources are preferentially located near the star clusters.

The distributions of the spatial displacements are similar in all three
galaxies, see Fig.~\ref{dhist}.  We compared the various distributions
using a Kolmogorov-Smirnov (KS) test and also by calculating the
average source displacement.  We applied both tests to both the full
data sample and also restricting the source luminosities to the range
$10^{36} {\rm \, erg \, s^{-1}} < L_X <  10^{38} {\rm \, erg \,
s^{-1}}$.  In all cases, the distributions of the spatial displacements
for the X-ray sources in the three galaxies are consistent with being
drawn from the same distribution.  The median displacement for NGC 1569
appears somewhat smaller than for the other galaxies, but the
difference is not statistically significant.  The 90\% confidence level
error intervals for the median overlap for all three galaxies.


\begin{figure} \psfig{file=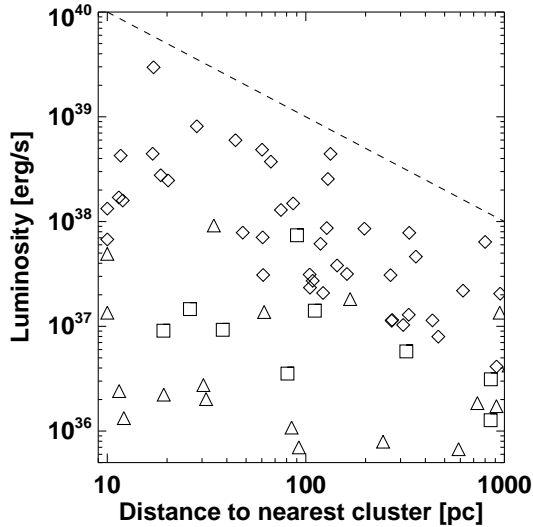,width=3.0in} \caption{X-ray source
luminosity versus displacement from nearest star cluster.  Sources from
M82 are shown as diamonds, NGC 1569 as triangles, and NGC 5253 as
squares.  The luminosities are for the 0.3--8~keV band.  The dashed
line represents the equation $L_x = (1 \times 10^{41} \, {\rm erg \,
s^{-1}}) / (d/{\rm pc})$.} \label{lumd} \end{figure}

Fig.~\ref{lumd} shows the luminosity of each X-ray source plotted
versus its spatial displacement from the nearest star cluster.   The
striking feature of the plot is that there are no high luminosity
sources at large displacements from star clusters.  Furthermore, there
is an apparent trend of decreasing luminosity with increasing
displacement from the nearest star cluster.  

To evaluate if this trend is statistically significant, we performed a
`boot-strap' analysis.  We used the set of displacements and
luminosities shown in Fig.~\ref{lumd} and then randomly re-arranged the
pairings.  This corresponds to the null hypothesis in which luminosity
and displacement are unrelated.  In $10^7$ trials, we found 1920 cases
in which no sources lay in the same region of the
luminosity-displacement plot which is empty in Fig.~\ref{lumd}.  This
corresponds to a chance probability of occurrence of $1.9 \times
10^{-4}$.

For M82 there is also an absence of dim sources at small
displacements.  This is likely due to the high level of diffuse X-ray
emission in the central 200~pc of M82 \cite{griffiths00} which
precludes detection of dim sources in the central regions of M82.  Dim
sources are detected very near star clusters in NGC 1569.

\section{Discussion}

We have shown that X-ray sources in these three starburst galaxies are
preferentially located near star clusters.  Because the star clusters
are very good tracers of current star formation activity in the
galaxies, this confirms that the X-ray binaries are young objects
associated with current star formation.  

We also found significant displacements of the X-ray sources from the
clusters.  Because X-ray binaries, unlike other bright X-ray sources
such as young supernovae, can exhibit high velocities, this suggests
that much of the X-ray source population consists of X-ray binaries.
Several mechanisms can put a binary in motion.  For neutron star
binaries, a `kick' due to an asymmetric explosion in the formation of
the neutron star can lead to high velocities \cite{lyne94}.  Even in
the  absence of `kicks' from supernova explosions, momentum
conservation following a symmetric ejection of matter in the formation
of the neutron star or black hole in a binary with a high mass
companion can produce a runaway speed of $\sim 50 \rm \, km \, s^{-1}$;
the ejected matter continues to move with the instantaneous orbital
velocity of the compact object at the moment of ejection and the binary
must move in the opposite direction to conserve momentum
\cite{nelemans99,vdh00}. For binaries in clusters, interactions with
other stars and binaries in the cluster can eject the binary from the
cluster \cite{phinney91,kulkarni93,sigurdsson93}. In young, dense star
clusters such interactions can occur on time scales of a few Myr
\cite{portegies99}.  Objects ejected via dynamical interactions tend to
escape with close to the minimum energy needed to escape
\cite{joshi01}.  The runaway velocities (at infinity) should be of the
same magnitude as the stellar velocity dispersions of the clusters,
which are typically $10-15 \, \rm km \, s^{-1}$ in these galaxies
\cite{smith01,mccrady03}.  The displacements we observe in the
starburst galaxies are likely due to motion of the X-ray sources caused
by one or more of these mechanisms.

Furthermore, we found that there is an absence of bright X-ray sources
with large displacements.  This suggests that there is some correlation
between the maximum possible brightness of an X-ray source and its
motion.  This correlation appears to hold only for (isotropic
equivalent) luminosities above $10^{38} \rm \, erg \, s^{-1}$.   The
excluded region appears to be bounded by a linear relation between
(isotropic equivalent) source luminosity $L_X$ and source displacement
from the nearest star cluster $d$, $L_x < (1 \times 10^{41} \, {\rm erg
\, s^{-1}}) / (d/{\rm pc})$.

In discussing this correlation, we first consider the case where the
X-ray sources emit isotropically.  In this case, the systems producing
such high luminosities likely contain black holes accreting via Roche
lobe overflow because such high luminosities would be difficult to
achieve in a wind accretor due to the low efficiency of wind capture
and black holes and needed to not violate the Eddington limit
\cite{blondin91,petterson78}. 

If the X-ray sources are ejected from the star clusters with speeds
which are roughly independent of mass, then the inverse correlation
between maximum X-ray source luminosity and displacement from the
nearest star cluster would arise if the source lifetime varies
inversely with luminosity.  An upper bound on the source lifetime can
be obtained from the time required to fully accrete the stellar
companion.  For a companion mass $M$ and an efficiency for the
conversion of mass lost by the companion to luminosity of $\eta$, the
source lifetime must be $T \le \eta M c^2 / L$ where $L$ is the average
luminosity and $c$ is the speed of light.  For sources traveling with a
speed $v$ perpendicular to the line of sight, the displacement from the
point of origin will then be $d \le v \eta M c^2 / L$.  If the
companion mass is independent of the compact object mass, then this
would reproduce the required dependence of source lifetime on
luminosity.  

Given a typical runaway velocity $v \sim 10 \, \rm km \, s^{-1}$, we
must have $\eta M \sim 0.2 M_{\odot}$ to match the line plotted in
Fig.~\ref{lumd} which bounds the region where X-ray sources are found. 
If Roche lobe overflow is occurring, then accretion onto the compact
object may be efficient with little mass loss giving $\eta \sim 0.1$. 
In this case, the companion mass would be $M \sim 2 M_{\odot}$.  Such
intermediate mass companions could be captured via dynamical
interactions in the cluster.  However, the capture must be directly
into a Roche-lobe filling orbit or the binary must hardened into a
Roche-lobe filling orbit via successive interactions in order to begin
accretion promptly, since the evolutionary time scale of the companion
is long.  A better understanding of the IMF and the dynamical
interactions within the clusters is needed to determine if this
scenario is viable.  Even with accretion via Roche lobe overflow, the
efficiency $\eta$ may be less than $0.1$ since outflows are often
observed in X-ray binaries.  Super-Eddington mass transfer would also
produce $\eta < 0.1$.  If $\eta < 0.1$, then a higher companion mass
may be compatible with the data shown in Fig.~\ref{lumd}.

If the X-ray sources have high-mass companions and maximum speeds near
$\sim 50 \rm \, km \, s^{-1}$, then the absence of high luminosity
source at large displacements implies a limit on the X-ray emitting
lifetime of the sources.  We find no sources at luminosities above
$10^{38} \rm \, erg \, s^{-1}$ at displacements larger than 200~pc.  At
$50 \rm \, km \, s^{-1}$, this would imply that the lifetimes of these
luminous sources must be less than 4~Myr, corresponding to very massive
stars.  An alternative is that the luminosity of the sources decreases
with age.  This would require an evolutionary path for binaries which
produces an accretion rate which decreases with age.

The X-ray binaries may also be beamed
\cite{king01,kording02,kaaret03}.  If the X-ray sources are high-mass
systems with high velocities, then the observed correlation would imply
that beaming only occurs when the binaries are quite young.  King et
al.\ (2001) suggest that the ULXs in starburst galaxies are high mass
X-ray binaries with beamed X-ray emission in a phase of
thermal-timescale mass transfer.  The delay between the formation of
the black hole (and, presumably, the start of the binary's motion away
from it point of origin) and the onset of the thermal-timescale mass
transfer phase depends on the stellar evolution of the companion.  The
delay could be $\sim$~20~Myr for a $9 M_{\odot}$ companion, which would
imply that very bright X-ray sources should be visible out to 1~kpc.
The data appear inconsistent with this, unless highly beamed X-ray
emission occurs in the thermal-timescale mass transfer phase only for
very massive companions.  Large displacements of high flux sources,
inconsistent with the data, also appear allowed in the relativistic
beamed model of Kording et al.\ (2002).


\section*{Acknowledgments}

We thank the Aspen Center for Physics for its hospitality during the
workshop where this work was begun.  PK acknowledges partial support
from NASA grant NAG5-7405 and Chandra grant GO2-3102X.  JSG thanks the
University of Wisconsin-Madison for support of this research.


\label{lastpage}


\begin{thebibliography}{99}


\bibitem[Alonso-Herrero et al.\ 2003]{alonso03} Alonso-Herrero A.,
Rieke G.H., Rieke M.J., Kelly D.M.\ 2003, AJ, 125, 1210

\bibitem[Beck et al.\ 1996]{beck96} Beck S.C., Turner J.L., Ho P.T.P.,
Lacy J.H., Kelly D.M.\ 1996, ApJ, 457, 610

\bibitem[Blondin, Stevens, \& Kallman 1991]{blondin91} Blondin J.M.,
Stevens I.R., Kallman, T.R.\ 1991, ApJ, 371, 684

\bibitem[Calzetti et al.\ 1997]{calzetti97} Calzetti, D.\ et al.\ 1997,
AJ, 114, 1834

\bibitem[de Grijs, O'Connell, \& Gallagher 2001]{degrijs01}  de Grijs
R., O'Connell R.W., Gallagher J.S.\ III 2001, AJ, 121, 768

\bibitem[de Grijs, Bastian, \& Lamers 2003]{degrijs03}  de Grijs R.,
Bastian N., Lamers H.J.G.L.M.\ 2003, MNRAS, 340, 197

\bibitem[Fabbiano 1989]{fabbiano89} Fabbiano G.\ 1989, ARA\&A, 27, 87

\bibitem[Fabbiano et al.\ 2003]{fabbiano03} Fabbiano G., Zezas A., King
A.R., Ponman T.J., Rots A., Schweizer F.\ 2003, ApJ, 584, L5

\bibitem[Gibson et al.\ 2000]{gibson00} Gibson, B.K.\ et al.\ 2000,
ApJ, 529, 723

\bibitem[Griffiths et al.\ 2000]{griffiths00} Griffiths, R.E.\ et al.\
2000, Science, 290, 1325

\bibitem[Hunter et al.\ 2000]{hunter00} Hunter D.A., O'Connell R.W., 
Gallagher J.S., Smecker-Hane T.A.\ 2000, ApJ, 120, 2383

\bibitem[Joshi, Nave, \& Rasio 2001]{joshi01} Joshi K.J., Nave C.P., 
Rasio F.A.\ 2001, ApJ, 550, 691

\bibitem[Kaaret 2001]{kaaret01b} Kaaret P.\ 2001, ApJ, 560, 715

\bibitem[Kaaret et al.\ 2001]{kaaret01} Kaaret P.\ et al.\ 2001, MNRAS,
321, L29

\bibitem[Kaaret et al.\ 2003]{kaaret03} Kaaret P., Corbel S., Prestwich
A.H., Zezas A.\ 2003, Science,  299, 365.

\bibitem[Kilgard et al.\ 2002]{kilgard02} Kilgard R.E., Kaaret P.,
Krauss M.I., Prestwich A.H., Raley M.T., Zezas A.\ 2002, ApJ, 573, 138

\bibitem[King et al.\ 2001]{king01} King A.R., Davies M.B., Ward M.J.,
Fabbiano G., Elvis M.\ 2001, ApJ, 552, L109

\bibitem[K\"ording et al.\ 2002]{kording02} K\"ording E., Falcke H.,
Markoff S.\ 2002, A\&A, 382, L13

\bibitem[Kulkarni, Hut, \& McMillian 1993]{kulkarni93} Kulkarni S.R.,
Hut P., McMillian S.\ 1993, Nature, 364, 421

\bibitem[Lyne \& Lorimer 1994]{lyne94} Lyne A.G., Lorimer D.R. 1994,
Nature, 369, 127

\bibitem[Makishima et~al.\ 2000]{makishima00} Makishima K.\ et~al.\
2000, ApJ, 535, 632

\bibitem[Martin, Kobulnicky, Heckman 2002]{martin02} Martin C.L.,
Kobulnicky H.A., Heckman T.M.\ 2002, ApJ, 574, 663

\bibitem[Massey \& Hunter 1998]{massey98} Massey P.M., Hunter D.A.\
1998, ApJ, 493, 180

\bibitem[McCrady, Gilbert, \& Graham 2003]{mccrady03} McCrady N.,
Gilbert A.M., Graham J.R.\ 2003, ApJ, 596, 240

\bibitem[McLeod et al.\ 1993]{mcleod93} McLeod K.K., Rieke G.H., Rieke
M.J., Kelly D.M.\ 1993, ApJ, 412, 111

\bibitem[Mengel et al.\ 2002]{mengel02} Mengel S., Lehnert M.D., Thatte
N., Genzel R.\ 2002, A\&A, 383, 137

\bibitem[Meurer et al.\ 1995]{meurer95} Meurer, G., Heckman, T.M.,
Leitherer, C., Kinney, A., Robert, C., Garnett, D.R.\ 1995, AJ, 110,
2665

\bibitem[Nelemans, Tauris, \& van den Heuvel 1999]{nelemans99} Nelemans
G., Tauris T.M., van den Heuvel E.P.J.\ 1999, A\&A, 352, L87

\bibitem[Petterson 1978]{petterson78} Petterson J.A.\ 1978, ApJ, 224,
625

\bibitem[Phinney \& Sigurdsson 1991]{phinney91} Phinney E.S.,
Sigurdsson S.\ 1991, Nature, 349, 220

\bibitem[Portegies Zwart et al.\ 1999]{portegies99} Portegies Zwart
S.F., Makino J., McMillian S.L.W., Hut P.\ 1999, A\&A, 348, 117

\bibitem[Prestwich et al.\ 2003]{prestwich03} Prestwich A.H.\ et al.\
2003, ApJ, 595, 719

\bibitem[Satyapal et al.\ 1997]{satyapal97} Satyapal S.\ et al.\ 1997,
ApJ, 483, 148

\bibitem[Smith \& Gallagher 2001]{smith01} Smith L.J., Gallagher J.S.
III 2000, MNRAS, 326, 1027

\bibitem[Sigurdsson \& Hernquist 1993]{sigurdsson93} Sigurdsson S.,
Hernquist L.\ 1993, Nature, 364, 423

\bibitem[Sternberg 1998]{sternberg98} Sternberg A.\ 1998, ApJ, 506, 721

\bibitem[Tremonti et al.\ 2001]{tremonti01} Tremonti C.A., Calzetti,
D., Leitherer, C., Heckman, T.M.\ 2001, ApJ, 555, 322

\bibitem[Turner, Beck, \& Ho 2000]{turner00} Turner J.L., Beck S.C., Ho
P.T.P.\ 2000, ApJ, 532, L109

\bibitem[van den Heuvel et~al.\ 2000]{vdh00} van den Heuvel E.P.J.,
Portegies Zwart S.F., Bhattacharya D., Kaper L.\ 2000, A\&A, 364, 563

\bibitem[Zezas \& Fabbiano 2002]{zezas02} Zezas A., Fabbiano G.\ 2002,
ApJ, 577, 726

\end{thebibliography}
\end{document}